\documentclass[aip,jcp,amsmath,amssymb,
preprint
]{revtex4-1}
\usepackage{graphicx}
\usepackage{dcolumn}
\usepackage{bm}
\usepackage[utf8]{inputenc}
\usepackage[T1]{fontenc}
\usepackage[normalem]{ulem}
\usepackage{xspace}
\usepackage[makeroom]{cancel}

\usepackage{color}

\newcommand{\elm}[3]{\left<#1\middle|#2\middle|#3\right>}
\newcommand{\scal}[2]{\left<#1\middle|#2\right>}

\definecolor{amethyst}{rgb}{0.6, 0.4, 0.8}
\definecolor{applegreen}{rgb}{0.55, 0.71, 0.0}
\definecolor{verde}{rgb}{0.0, 0.5, 0.0}
\definecolor{arjan}{rgb}{0.6, 0.2, 0.4}

\begin{document}

\title[TPS and polarizability]{The localization spread and polarizability 
of rings and 
periodic chains}

\author{Celestino Angeli}
\email{anc@unife.it}
\affiliation{Dipartimento di Scienze Chimiche, Farmaceutiche ed Agrarie, Universit\`a di Ferrara, via Borsari 46, 44121 Ferrara, ITALY}

\author{Gian Luigi Bendazzoli}
\affiliation{Universit\`a di Bologna, Bologna, ITALY}

\author{Stefano Evangelisti}
\email{stefano.evangelisti@univ-tlse3.fr}
\affiliation{Laboratoire de Chimie et Physique Quantiques, UMR5626, Universit\'e de Toulouse
(UPS), CNRS, 118 Route de Narbonne, F-31062
Toulouse, France}

\author{J. Arjan Berger}
\affiliation{Laboratoire de Chimie et Physique Quantiques, UMR5626, Université de Toulouse (UPS), CNRS, and European Theoretical Spectroscopy Facility, 118 Route de Narbonne, F-31062
Toulouse, France}

\date{\today}

\begin{abstract}
The localization spread gives a criterion to decide between metallic versus insulating behaviour of a material. 
It is defined as the second moment cumulant of the many-body position operator, divided by the number of electrons. 
Different operators are used for systems treated with Open or Periodic Boundary Conditions.
In particular, in the case of periodic systems, we use the complex-position definition, that was already used in similar contexts for the treatment of both classical and quantum situations.
In this study, we show that the localization spread evaluated on a finite ring system of radius $R$ with Open Boundary Conditions leads, in the large $R$ limit, to the same formula derived by Resta {\it et al.} for 1D systems with periodic Born-vonK\'arm\'an boundary conditions. 
A second formula, alternative to the Resta's one, is also given, based on the sum-over-state formalism, allowing for an interesting generalization to polarizability and other similar quantities.\\
\end{abstract}

\maketitle

\newpage
\section{Introduction}

The position operator 
 $\hat {\bf r}$ plays a crucial role in Quantum Mechanics.
Indeed, it is very often the key element to build the potential operator.
Moreover, in a single-particle description, it is used to define 
multipole moments and polarizabilities.
Finally, its spread is one of the key ingredients that 
enter the Heisenberg Uncertainty Principle.
A similar crucial role occurs in many-particle systems.
In this case, the one-body position operator $\hat{\bf r}_\mu $ of each
particle $\mu$ can be combined in order to give the total-position
operator:
\begin{equation}
    \hat{\bf Q}=\sum_\mu \hat{\bf r}_\mu.
\label{totalp}
\end{equation}
This operator is by definition a quantity that refers to the entire system as a whole.
In a series of papers, Resta and co-workers\cite{Resta98,Resta99,Resta02,Resta06} and then Souza \textit{et al},\cite{Souza00} 
 after an original idea that goes back to Kohn more than
fifty years ago,\cite{Kohn64}
showed that the spread of the total position,
called by the authors  {\it Localization Tensor} once it is divided by the number of
identical particles, is able to discriminate between systems that behave as
insulators or conductors in the thermodynamic limit.
Indeed, the {\em per-electron} position spread (\textit{i.e.}, the localization tensor)
diverges in the case of metals, while it remains finite for insulators.
Some of us have recently used the localization tensor to study the Wigner
localization.\cite{Diaz-Marquez_2018,Brooke2018, Escobar}
However, it has been shown that in some cases border effects can play a very important role 
and completely hide the behavior of the rest of the system.\cite{Evang2010,Monari-Evang}
For this reason, the extension of these ideas to periodic systems has attracted much attention. \cite{Otto92,Kudino91,Kudino99a,Kudino99b}

In Quantum Mechanics, the spread of any operator $\hat A$ is given by the standard expression
\begin{equation}
    \bar{\bar A} \; = \; \elm{\Psi }{\hat A^2}{\Psi} \, - \,  \elm{\Psi }{\hat A}{\Psi}^2.
\label{lambdaopenbc}
\end{equation}
When $\hat{A}=\hat{\bf Q}$  we get the Total-Position Spread, denoted in the following as TPS.
Indeed, this is the way the position spread is computed for finite systems.
We systematically calculated the TPS for finite molecular systems, in which case this quantity gives interesting information on the nature of bonds and the mechanism of bond breaking.\cite{Angeli13,Brea13,Khatib14,Benda14,Khatib15,Huran16}
If the size of the system  is systematically increased, the thermodynamic limit can be computed by extrapolating finite calculations to the infinite-size limit.\cite{Vetere08,Bendazzoli_2008,Monari08,Vetere09,Benda11,Benda12,Benda15,Ferti15,Battag18,Diaz18,Huran18}

However, for practical reasons, very large (``infinite'') systems are often described within the framework of {\em periodic}, or Born-von Karm\'an  boundary conditions (in this context), and this poses a subtle theoretical problem. 
Indeed, in the Periodic Boundary Condition (PBC) formalism, the position operator is not a single-valued function, because an infinite set of values of the periodic coordinates correspond to the same point in the system.
For this reason, the position spread for periodic systems must be defined in a different way.
\\

The problem was addressed by Resta {\em et al.}, in the context of the
so called modern theory of polarization.\cite{Resta06} The central quantity is $\hat{U}$, 
the exponential of the total position defined in Eq. (\ref{totalp}) which is a $N-$body operator
and it is used to define the localization spread $\lambda_R$.
In case of a 1D system of $N$ electrons and length $L$, one has:
\begin{eqnarray}
  \hat{U} & = & \exp \left( \frac{2 \pi i}{L} \sum_{j=1}^N x_j\right), \nonumber \\
  \lambda_R & = & - \frac{L^2}{4 \pi^2 N} \ln|\langle \Psi | \hat{U} | \Psi \rangle |^2.
  \label{lambdaR}
\end{eqnarray}
Later, C. Sgiarovello {\it et al.} derived a formula for the computation of the thermodynamic limit of Eq. (\ref{lambdaR}) for 
a determinantal wavefunction and applied it to some crystalline systems. \cite{sgpere}
\\ 

We recently addressed this problem by adopting a different strategy.~\cite{Valenca}
We notice that all functions of the position that have the same periodicity of
the whole system are perfectly acceptable quantities.
This is the case, for instance, for the periodic potentials defined for this type of systems. Our approach (see Refs. [\onlinecite{Escobar},\onlinecite{Valenca}])
is to redefine the one-particle position operator itself, essentially replacing
the position by the imaginary exponent of the position.
In doing that, one must assure two basic requirements:
\begin{enumerate}
    \item The new operator must have the same periodicity as the PBC system.
    \item The difference between two operators corresponding
    to fixed values of the coordinates must 
    tend, in the limit of infinite system and up to a phase factor, 
    to the corresponding difference obtained from the ordinary position operator.
\end{enumerate}
The above conditions can be satisfied in different ways.
In our previous work (Ref. [\onlinecite{Valenca}]), we defined a complex position operator as
\begin{equation}
    {\hat q}_L (x) \; = \; \frac{L}{2\pi i} \left[e^{\frac{2\pi i }{L}x}-1\right].
    \label{complexpositionfirst}
\end{equation}
This choice has the advantage that ${\hat q}_L (x)$  reduces to the standard 
position operator when $x/L \ll 1$, \textit{i.e.} ${\hat q}_L (x)=x$.
In the present context, we compute a cumulant of the square norm of the position.
Because of this fact, the constant shift $-\frac{L}{2\pi}$  in Eq. (\ref{complexpositionfirst})
can be dropped, as well as the imaginary unit.
In case of a 1D system of
length $L$, we can simply use the quantity:
\begin{equation}
    {\hat q}_x \; = \; \frac{L}{2\pi} e^{\frac{i2\pi x}{L}}.
    \label{complexposition}
\end{equation}
%
%
We notice that this definition of the position is not restricted to the Quantum-Mechanics context.
Indeed, it has been used in Classical Physics, in order to perform Madelung sums for ionic
systems,\cite{Tavernier_2020,Tavernier_2021} and to compute the classical energy and harmonic and anharmonic corrections of Wigner Crystals.\cite{Alves21}
In Appendix III, a detailed discussion on the choice of the position operator for periodic systems is presented.
\\

In this paper we assume a slightly different starting point. We
consider the localization spread of a ring system with the { \em open boundary
conditions} (OBC) where the definition  of Eq. (\ref{lambdaopenbc}) holds, and we obtain the same
results one gets with the {\em complex position operator } of Eq. (\ref{complexposition}) for a periodic system. 
Moreover, we also get the formula of Ref. [\onlinecite{sgpere}], which was derived from the formalism of Resta.
In detail, we can summarize the scheme of the present paper as follows, which is
concerned with rings with OBC and 1D systems with PBC:
 we first derive formulae for the TPS and the polarizability of a one-determinant 
wavefunction of many electrons  in a ring under a potential of $C_n$ symmetry; then, thanks to the isomorphism of $C_n$ and the translation in a 1D system with Born-von Karm\'an PBC, all the treatment extends to the latter; the formula for the TPS shows that a partially filled band leads to a per electron TPS diverging in the thermodynamic limit; the formula for the TPS is alternative but equivalent to the Sgiarovello-Peressi-Resta\cite{sgpere} one for a complete orbital basis; finally, we show applications to the H\"uckel  wavefunction for dimerized annulene and cyclacene, where closed analytical solutions are found. This approach is called 
tight-binding (TB) in the physical literature.
 
For the sake of simplicity, as previously said, we will focus on one dimension in the whole of this paper and the generalization to higher dimensions will be addressed in forthcoming papers. Finally, we stress the fact that atomic units (bohr, hartree, etc.) will be used in the whole of the presentation.

\section{Particles in a ring under a periodic potential ($C_n$).}
\subsection{General considerations}
\label{sec:bloch}
Let us consider a system of non interacting electrons moving in a ring of length $L$ and radius $R=L/2\pi$ and subject to a non constant potential $U$ of  $C_n$ symmetry. 
Its wavefunction will be a Slater determinant of spinorbitals that can be taken to be eigenfunctions of $\hat{C}_n$, the 
anticlockwise rotation of $2 \pi/n$ around the centre of the ring. The structure of such orbitals 
is that of Bloch orbitals for 1D periodic systems (see the Supplementary material for details). This is due to the isomorphism of the 
$C_n$ group generated by the in-plane rotation $\hat{C}_n$ of an angle $ 2 \pi /n$ and the 
group $T_n$ generated by the translation $\hat{t}_d$ of a displacement $d$  when 
acting on the space of periodic functions of period $L=nd$, according to  
the Born-vonK\'arm\'an boundary conditions. Actually, these two groups are both examples of finite cyclic groups and this is the reason of the isomorphism.\cite{grouptheory,Calais_1995}

The eigenfunctions of $\hat{C}_n$ have the following Bloch structure:
\begin{equation}
 \psi(s) \; = \; \psi_k(s) \; = \; e^{ \frac{2 \pi i k s}{L}} \: u_k(s) \; = \;
                                   e^{i K s} \: u(s,K),
\label{bloch}
\end{equation}
where $u_k(s)=u_k(s+d)$ is a periodic function and $k$ is an integer defined $\mod n$. In order to conform to the 
solid-state literature we introduced the (discrete) variable $K \: = \: \frac{2 \pi k }{L}$
and the alternative notation $u(s,K)$ for $u_k(s)$.
The structure of the function given in Eq. (\ref{bloch}) can be described as 
a plane wave modulated by a periodic factor $u(s,K)$. The discrete variable $K$ 
becomes (quasi-)continuous for large $n$.

The proper definitions of the orbitals taking into account 
normalization are, in the two notations:
\begin{eqnarray}
\mbox{integer} \; k:  \quad  
\psi_k(s) & = & \frac{1}{\sqrt{n}} e^{\frac{ 2 \pi i k s}{nd}} u_k(s),  
\label{ukpiccolo} \\
 K = \frac{2 \pi k}{nd}:  \quad  
\psi(s,K) & = & \frac{1}{\sqrt{n}} \, e^{iKs} \, u(s,K). 
\label{ukgrande}
\end{eqnarray}

\subsection{Approximate wavefunctions.}
\label{approxwf}

Exact solutions of the Schr\"odinger equation with a periodic Hamiltonian
are known only in exceptional cases and in practice one resorts to 
variational treatments by expanding the orbitals in suitably chosen basis 
functions, like in the well known LCAO approximation.  
We  place in 
each cell $\mu$ a number $n_c$ of basis functions $\chi_j(s), \, j=1,2, \ldots, n_c$ centered 
in $n_c$ points $s_{1\mu}, \, s_{2\mu}, \, \ldots, s_{n_c \mu}$, 
$s_{j\mu} \,= \,s_{j0}+\mu d$. We introduce the symmetry-adapted basis functions:
\begin{equation}
\begin{array}{ccccl}
b_{kj}(s) & = & \sum_{\mu=0}^{n-1} 
   e^{\frac{ 2 \pi i k \mu}{n}} \chi_j(s-s_{j \mu}) 
& = & e^{i K s} \sum_{\mu=0}^{n-1} 
      e^{\frac{ 2 \pi i k (\mu d -s)}{nd}} \chi_j(s-s_{j \mu}), \; 
j=1,2, \ldots, n_c 
 \\
\hat{C}_n \, b_{kj}(s) & = & \sum_{\mu=0}^{n-1} 
  e^{\frac{ 2 \pi i k \mu}{n}} \chi_j(s-s_{j \mu+1})  
 & = &  e^{- \frac{ 2 \pi i k }{n}} b_{kj}(s) 
\end{array} 
\label{bimu}
\end{equation}
The total number of the $b_{kj}$'s is $n \times n_c$. 
\\ 

The matrix elements  of the overlap $S$ and of the hamiltonian $\hat{H}$ 
%
%
in the symmetry adapted basis are: 
\begin{eqnarray}
\scal{b_{kj}}{b_{k'j'}} & = & \delta_{kk'} \,
\sum_{\mu \mu'} \, e^{\frac{ 2 \pi i k (\mu'-\mu)}{n}}
\scal{\chi_{j \mu}}{\chi_{ j' \mu'}}\!\!, \\
\label{ovlvar}
\elm{b_{kj}}{\hat{H}}{b_{k'j'}\!} & = & \delta_{kk'} \!\!\!
\sum_{\mu \mu'} \! e^{\frac{ 2 \pi i k (\mu'-\mu)}{n}} 
\!\elm{\chi_{j \mu}}{\hat{H}}{\chi_{ j' \mu'}\!}\!\!. 
\label{hamvar}
\end{eqnarray}
Given that $[\hat{H},\hat{C}_n]=0$, the  matrix of $\hat{H}$ assumes a block structure: 
there are $n$ blocks $\mathbf{H}_k$ and $\mathbf{S}_k$ each of dimension 
$n_c \times  n_c$  that can be diagonalized to get the variational solution.
If $c_{1 \gamma}, \, c_{2 \gamma}, \ldots, c_{n_c \gamma}$ is the $\gamma-$th 
eigenvector of $\mathbf{H}_k$ in the metric $\mathbf{S}_k$, one has the variational 
solution
\begin{eqnarray}
	\psi_{\gamma k}(s) & = & \mathcal{N} \; \sum_{j=1}^{n_c} \,c_{j \gamma} b_{kj}(s)
	\; = \; \mathcal{N} \sum_{\mu=0}^{n-1} \; e^{\frac{ 2 \pi i k \mu}{n} } 
       \,\sum_{j=1}^{n_c} \,c_{j \gamma}  \chi_j(s-s_{j \mu}), 
\label{varsol1}
\end{eqnarray}
where $\mathcal{N}$ is the normalization constant. The wavefunction in Eq. (\ref{varsol1}) can
be rewritten in the form reported in Eq. (\ref{ukgrande}) with its periodic factor defined as follows: 
\begin{eqnarray}
u_\gamma(s,K) \! & = & \! \sum_{\mu=0}^{n-1} e^{\frac{2 \pi i k (\mu d-s)}{nd}}
  \sum_j\! c_{j \gamma} (K) \chi_j(s-s_{j \mu}). 
\label{uvariational}
\end{eqnarray} 
%
%
\\

From Eq. (\ref{hamvar}) one finds that the blocks $\mathbf{H}_k$ and
$\mathbf{H}_{-k}$ are complex conjugated, but both are hermitean matrices
so their eigenvalues are the same. The corresponding eigenfunctions 
can be grouped in couples with the same energy and behave like  
degenerate eigenvectors belonging to a 2-dimensional IR of a non abelian
group.
Besides the variational treatment, further approximations may be adopted
to simplify the computation of  the matrix elements of the hamiltonian matrix.
As a limit case of such an approach we may consider the well known H\"uckel model.
The expansion basis are site functions $\chi$ centered in a point $P_{j \mu}$ 
and are supposed to be orthonormal eigenfunctions
of the position operators. This is the common practice although these site
functions are rather awkward mathematical objects, see \textit{e.g.} Ref. [\onlinecite{sqrdelta}].
Accordingly, the $\chi$'s are everywhere vanishing
but in $P$. As concerns the hamiltonian matrix elements this basis, 
they are treated as adjustable parameters 
assumed to be zero except for $\chi$ functions placed on
nearest neighbour sites. In solid state physics such hamiltonian parameters
are known as {\it hopping integrals} and denoted by the symbol $t$, while
in quantum chemistry the name {\it resonance integral} and the symbol
$\beta=-t$ are preferred. The advantage of the H\"uckel model is its exact 
solubility in a number of cases, combined with an ability to gain insight
into the electronic structure and properties.\cite{kutzel} This is the reason why the 
examples we provide are concerned with H\"uckel wavefunctions.

\subsection{The TPS of \textit{n} electrons in a ring.}

We now consider a $n$-electron determinantal  wave function $\Phi$ constructed using the 
Bloch orbitals defined in Eq. (\ref{bloch}) and the total position operators
\begin{equation}
\hat{X} \: = \: \sum_{j=1}^n \: x_j, \quad \hat{Y} \: = \: \sum_{j=1}^n y_j.
\end{equation}
The TPS tensor $\mathbf \Lambda$ of a ring is diagonal and its $xx$ and $yy$ 
components are equal; \footnote{strictly this holds for $n>2$} for this reason we 
may consider its trace: 
\begin{eqnarray}
Tr(\mathbf{\Lambda})  & = & \mathbf{\Lambda}_{xx} + \mathbf{\Lambda}_{yy}
\nonumber \\
 & = & 
 \elm{\Phi}{\hat{X} \hat{X} \! + \! \hat{Y}\hat{Y}}{\Phi} -
 \elm{\Phi}{\hat{X}}{\Phi}^2 - 
 \elm{\Phi}{\hat{Y}}{\Phi}^2 
\nonumber \\
\quad & = & \elm{\Phi}{(\hat{X} \pm i \hat{Y}) (\hat{X} \mp i \hat{Y})}{\Phi}. 
\label{xpiuiy}
\end{eqnarray} 

In Eq. (\ref{xpiuiy}) we introduced the operators $\hat{X} \pm i \hat{Y}$ in order to 
take advantage of the $C_n$ symmetry of the system, which ensures that $\elm{\Phi}{\hat{X}}{\Phi}=\elm{\Phi}{\hat{Y}}{\Phi}=0$ 
and $\elm{\Phi}{\hat{X}\hat{Y}}{\Phi}=0$,
\ref{bloch}) and reminding that $R = L / 2 \pi$, we find
One can show (see the Supplementary material) that the operator $x \pm i y$ shifts
by one unit the value of $k$ associated to a Bloch orbital:
\begin{eqnarray}
 (x \pm i y) \psi_{k \gamma}(s) & = & R\left[\cos\left(\frac{s}{R}\right) \pm
 i \sin\left(\frac{s}{R}\right) \right]
          \! e^{ \frac{ 2 \pi i k s }{ L}} \! u_{k \gamma}(s) \nonumber \\
\quad & = & R \, e^{\pm \frac{ i s }{ R} } 
          e^{\frac{ 2 \pi i k s }{ L} } \, u_{k \gamma}(s) \nonumber \\
\quad & = & R \,e^{\frac{ 2 \pi i (k \pm 1) s }{ L}} \, u_{k \gamma}(s).
\label{xpiuiy_bis}
\end{eqnarray}
where $s$ is the arc length.
More interesting, Eq. (\ref{xpiuiy_bis}) shows that on a circle of length $L$ one has: 
\begin{equation}
x \pm i y = \frac{L}{2 \pi} e^{ \pm \frac{2 \pi i s }{L}}.
    \label{x+iy}
\end{equation}
This quantity is nothing but the complex position operator defined in Eq. (\ref{complexposition}) for a periodic system
of period $L$ where $s$ is the ordinary position. 
Consequently, the results obtained in the sequel for a ring with OBC can be transferred to a 1D system with PBC. Eq. (\ref{x+iy}) provides a new interpretation of the complex position operator defined in Eq. (\ref{complexposition}).
\\

The function $\tilde{\psi}_{k \pm 1, \gamma} \; = \; (x \pm i y) \psi_{k \gamma}$ 
will not be in general eigenfunction
of $\hat{H}$ because of the mismatch between the quantum number $k$ of 
$u_{k \gamma}(s)$ and that of the associated plane wave. 
However, $\tilde{\psi}_{k \pm 1, \gamma}$ is still eigenfunction of $\hat{C}_n$ 
because it keeps the structure of Eq.  (\ref{bloch}).
The one-electron matrix elements of $x \pm i y$ are given by:
\begin{equation}
\elm{\psi_{k \gamma} }{x \pm i y}{\psi_{k' \gamma'}}  =
\!\! R \delta_{k,k' \mp1} \!\! \int_{0}^{d} \!\! u_{k, \gamma}(s)^*  u_{k \mp 1,\gamma'}  ds.
\label{matrel}
\end{equation} 
\\

The operators  $ \hat{X} \pm i\hat{Y}$ transform a Slater determinant $\Phi$ into a sum
of single excitations, by replacing each occupied spin orbital $\psi_{k \gamma \sigma}$
with $\tilde{\psi}_{k \pm 1, \gamma \sigma}$. In order to simplify the notation we
introduce a multi-index $j={k \gamma \sigma}$ to address the 
spin orbital $\psi_{k \gamma \sigma}$ and $\tilde{\jmath}$ for the spin orbital
$(x \pm i y) \psi_{k \gamma \sigma}$:
\begin{equation}
(\hat{X}\pm i\hat{Y}) \, \Phi \; = \; \sum_{j} \Phi_{j}^{\tilde{\jmath}} .
\label{singles}
\end{equation}
In Eq. (\ref{singles}) multi-indexes $j, \, \tilde{\jmath}$ span the occupied spin orbitals
and $\Phi_{j}^{\tilde{\jmath}}$ denotes the single excitation 
$\psi_j \rightarrow (x+iy) \psi_j$. By noticing that 
$\scal{\Phi}{\Phi_{j}^{\tilde{\jmath}}} \: = \: \ \elm{\psi_{k \gamma}}{x \pm i y}{\psi_{k \gamma}} \; = \;0 $ because of Eq. (\ref{matrel}), one has: 
\begin{equation}
\elm{\Phi}{ \hat{X} \pm i\hat{Y}}{ \Phi} = \; 0.
\label{XYzero}
\end{equation} 
Indeed, each determinant is eigenfunction  of $\hat{C}_n$ and its eigenvalue is the sum 
of the $k$ quantum numbers of the occupied spin-orbitals. Accordingly, all excitations in Eq. (\ref{singles}) differ by one unit in $k$ from $\Phi$ and Eq.
(\ref{XYzero}) follows.
\\
By using the result of Eq. (\ref{XYzero}), Eq. (\ref{xpiuiy}) can be written as:
\begin{eqnarray}
 Tr(\mathbf{\Lambda}) & = & \elm{ \Phi }{ \left(\hat{X} \pm i \hat{Y}\right)\left( \hat{X} \mp i \hat{Y}\right) }{ \Phi} 
 \;=\; \scal{ \sum_{j} \Phi_{j}^{\tilde{\jmath}} }{ \sum_{j'} \Phi_{j'}^{\tilde{\jmath}'}},
\label{trace}
\end{eqnarray}
where $j, \,j'$ span the occupied spin orbitals. To compute Eq. (\ref{trace}) we consider two possibilities:
\begin{enumerate}
\item we compute Eq. (\ref{trace}) directly involving only occupied orbitals;
\item sum over states: we expand each $ \sum_{j} \Phi_{j}^{\tilde{\jmath}}$ 
in the space spanned by the usual single excitation from occupied to virtual 
spin orbitals.
\end{enumerate}

\subsubsection{Direct computation.}
\label{sec:direct}

In order to compute Eq. (\ref{trace}) we use the following results: 
\begin{equation}
\scal{\Phi_{j}^{\tilde{\jmath}}}{\Phi_{j'}^{\tilde{\jmath'}}} \!\! = \!\! \left\{
\begin{array}{ll}
\!\! \scal{\tilde{\psi}_j}{\psi_j} \scal{\psi_{j'}}{\tilde{\psi}_{j'}} & 
 \mbox{if } j \ne j' \\
\quad & \quad \\
\!\!\scal{\tilde{\psi}_j}{\tilde{\psi}_j}\!- \!\! \sum_{m \ne j}\limits
\scal{\tilde{\psi}_j}{\psi_{m}} \scal{\psi_{m}}{\tilde{\psi}_j} &
\mbox{if } j = j'.
\end{array} \right.
\label{singsing}
\end{equation}
By noticing that $\langle \tilde{\psi}_j | \psi_j \rangle \; = \; 0$, because the two $\psi$'s
correspond to different eigenvalues of $C_n$, the double summation $\sum_{jj'}$ 
becomes $\sum_j$ and we find:
\begin{eqnarray}
 Tr(\mathbf{\Lambda}) & = & \sum_j \scal{\tilde{\psi}_j}{\tilde{\psi}_j} - \sum_{jm} \scal{\tilde{\psi}_j}{\psi_m} \scal{\psi_{m}}{\tilde{\psi}_j} \nonumber \\
& = &
\sum_j \scal{\tilde{\psi}_j}{\tilde{\psi}_j} - 
 \sum_{j} \elm{\tilde{\psi}_j}{\hat{P}_{occ}}{\tilde{\psi}_j},
\label{ltrace} 
\end{eqnarray} 
where $\hat{P}_{occ}\; = \; \sum_m | \psi_{m} \rangle \langle \psi_{m} | $ is the projection on the occupied  orbital 
subspace, because
the multi-indexes $j, \, m$ label the occupied spin orbitals. 
Eq. (\ref{ltrace}) separates in contributions from each spin 
$\sigma=\alpha \: \text{or} \: \beta$ as follows: 
\begin{eqnarray}
 Tr(\mathbf{\Lambda})_\sigma & = &  \left[ \sum_{\gamma k} \left(
 \scal{\tilde{\psi}_{k \gamma}}{\tilde{\psi}_{k \gamma}} -
 \elm{\tilde{\psi}_{k \gamma}}{\hat{P}_{occ}}{\tilde{\psi}_{k \gamma}} \right) \right]_\sigma \nonumber \\  
& = & \sum_{\gamma k} \, Tr(\mathbf{\Lambda})_{\gamma  k \sigma} ,
\label{preresta}
\end{eqnarray}
where only occupied orbitals of the given spin are involved in the sums. 
Eq. (\ref{preresta}) shows the contribution 
$Tr(\mathbf{\Lambda})_{\gamma  k \sigma}$ of each occupied spin orbital to
$Tr(\mathbf{\Lambda})$ and we notice that it cannot be negative because 
$1 - \hat{P}_{occ}$ is a projection.  Then we find:
\begin{eqnarray}
\scal{\tilde{\psi}_{k \gamma}}{\tilde{\psi}_{k \gamma}} & = &
R^2 \,\int_{0}^{d} \, u_{k \mp 1,\gamma}(s)^* \, u_{k \mp 1,\gamma}(s) \, ds \; \nonumber \\
& = & \; R^2 \; = \; 
\left( \frac{nd}{2 \pi} \right)^2 .
\end{eqnarray}
As concerns the 2nd term, 
$\elm{\tilde{\psi}_{k \gamma}}{\hat{P}_{occ}}{\tilde{\psi}_{k \gamma}}$, of Eq. (\ref{preresta}),
by taking into account Eq. (\ref{matrel}) it can be rewritten as follows:
\begin{eqnarray}
&&\!\!\!\elm{\tilde{\psi}_{k \gamma}}{1 - \hat{P}_{occ}}{\tilde{\psi}_{k \gamma}} = \nonumber \\
&&=
\scal{\tilde{\psi}_{k \gamma}}{\tilde{\psi}_{k \gamma}} -
\sum_{\gamma '} \scal{\tilde{\psi}_{k \gamma}}{\psi_{k \pm 1 \gamma '}} 
\scal{\psi_{k \pm 1 \gamma '}}{\tilde{\psi}_{k \gamma}} \nonumber  \\
   && =  R^2  \left( 1 - 
  \sum_{\gamma '} \left| \int_0^d u_{k \gamma}(s) \, u_{k \pm 1 \gamma '} (s) ds \right|^2 \right),
\label{direct}
\end{eqnarray} 
where all  indexes refer to {\it occupied orbitals} of the given spin. 
In this connection 
 we point out an essential difference between completely and partially filled bands. 
Consider a partially filled band up to a Fermi value $k_F$: the orbital 
$\tilde{\psi}_{k_F \gamma }$ will have zero projection in the occupied space 
of the band $\gamma'=\gamma$, while this is not the 
case in a completely filled band, because $k$ is defined $\mod n$. 
Therefore  $Tr(\mathbf{\Lambda})_{\gamma k_F \sigma}$ diverges for $n \rightarrow \infty$ 
as $R^2=O(n^2)$ and the localization per electron  
$Tr(\mathbf{\lambda})_\sigma=Tr(\mathbf{\Lambda})_\sigma/(n_cn)$ will diverge as $O(n)$
for $n \rightarrow \infty$. 
\\

Eq. (\ref{direct}) can be used to compute numerically $Tr(\mathbf{\lambda})_\sigma$ for a finite
system; in case of a partly filled band the sum $\sum_{\gamma '}$ 
is missing for some value of $k$ and $\gamma'=\gamma$.
As concerns the other values of $k$ and $\gamma'$, 
in order to compute the limit for $n \rightarrow \infty$ it is convenient to use the variable 
$K=2 \pi k / (nd) $ instead 
of $k$ and consider $u_{k,\gamma}(s)$ as a function of the continuous variable $K$:
\begin{eqnarray}
&& u_{k,\gamma}(s) \; \leftrightarrow \; u_{\gamma}(s,K), \nonumber \\  
&& u_{k \pm 1,\gamma}(s) \; \leftrightarrow \; u_{\gamma}(s,K \pm \Delta K) ,
\label{kcontinuo}
\end{eqnarray}
where $\Delta K \; =\; 2 \pi /(nd)$. Now, for large $n$ we write
\begin{eqnarray}
 u_{\gamma}(s,K \pm \Delta K) & = & u_{\gamma}(s,K)  \pm  
 \frac{2 \pi}{n d} \frac{\partial u_\gamma}{\partial K}  +  \nonumber \\
&&+ \frac{ 2 \pi^2}{n^2d^2}  \frac{\partial^2 u_\gamma}{\partial K^2} \; +
  O(n^{-3}) 
\label{utaylor}
\end{eqnarray}
and therefrom:
\begin{eqnarray}
&&\left| \scal{u_{k ,\gamma}}{u_{k \pm 1,\eta}}\right| ^2 =\nonumber \\
~~~&=&
\delta_{\gamma \eta} \pm 
\delta_{\gamma \eta}  
\frac{ 2 \pi}{nd} \left( 
\scal{u_{\gamma}}{\frac{\partial u_\eta}{\partial K}} + 
\scal{\frac{\partial u_\eta}{\partial K}}{u_{\gamma}} \right) + \nonumber \\ 
~~~&& +\left( \frac{ 2 \pi}{nd} \right)^2 
\scal{u_{\gamma}}{\frac{\partial u_{\eta}}{\partial K}} 
\scal{\frac{\partial u_{\eta}}{\partial K}}{u_{\gamma}}  + 
\nonumber \\
~~~&&+\frac{\delta_{\gamma \eta} }{2}\!\!
\left( \frac{ 2 \pi}{nd} \right)^2 \!\!
\left(  \scal{u_{\gamma}}{\frac{\partial^2 u_{\eta}}{\partial K^2}}+    
        \scal{\frac{\partial^2 u_{\eta}}{\partial K^2}}{u_\gamma} \right) +\nonumber \\
~~~    &&    + \; O(n^{-3}) \nonumber \\
~~~&=&  \delta_{\gamma \eta} \: + \: \left( \frac{ 2 \pi}{nd} \right)^2 
\scal{u_\gamma}{\frac{\partial u_\eta}{\partial K}}
\scal{\frac{\partial u_\eta}{\partial K}}{u_\gamma} + \; \nonumber \\
~~~&&- \delta_{\gamma \eta} \:
\left( \frac{ 2 \pi}{nd} \right)^2 
\scal{\frac{\partial u_\gamma}{\partial K}}{\frac{\partial u_\gamma}{\partial K}}
+ \; O(n^{-3}),
\label{intsquare}
\end{eqnarray}
where we used the relations
$ \frac{\partial}{\partial k} \scal{u_{K \gamma}}{u_{K' \gamma'}} \;= 0$
and $\frac{\partial^2}{\partial k^2} \scal{u_{K \gamma}}{u_{K' \gamma'}}\;= 0$.
For each value of $k$ such that $k \pm 1$ is occupied, Eq. (\ref{preresta})
involves the integrals  
$\scal{\psi_{k \mp 1 \, \gamma}}{\tilde{\psi}_{k \gamma'}}$ and 
from Eq. (\ref{intsquare}):
\begin{eqnarray}
Tr(\mathbf{\lambda})_{k \sigma} & = &\frac{1}{n} 
 \sum_{\gamma} \left\{ \! R^2\! -\! \sum_{\gamma'}  
\scal{\psi_{k \mp 1 \gamma}}{\tilde{\psi}_{k \gamma'}} 
\scal{\tilde{\psi}_{k \gamma'}}{\psi_{k \mp 1 \, \gamma}} \!\right\} \ \nonumber  \\
&=& \frac{R^2}{n}  \sum_{\gamma} \left\{ 1  -  \sum_{\gamma'} \left[
\delta_{\gamma \gamma'}  
+\right. \right.\nonumber \\& & 
+\left( \frac{2 \pi}{nd} \right)^2
\scal{u_\gamma}{\frac{\partial u_{\gamma'}}{\partial K}} 
\scal{\frac{\partial u_{\gamma'}}{\partial K}}{u_\gamma}+
\nonumber \\
&& - \left.\left.
\delta_{\gamma \gamma'} \left( \frac{2 \pi}{nd} \right)^2 
\scal{\frac{\partial u_{\gamma}}{\partial K}}{\frac{\partial u_{\gamma}}{\partial K}} 
\, + \, O(n^{-3}) \right] \right\} \nonumber \\
&=& \frac{1}{n} \left\{ \!\sum_{\gamma}
\scal{\frac{\partial u_{\gamma}}{\partial K}}{\frac{\partial u_{\gamma}}{\partial K}} + \right. \nonumber \\
&& - \left.\sum_{\gamma \gamma'}  
\scal{u_\gamma}{\frac{\partial u_{\gamma'}}{\partial K}} 
\scal{\frac{\partial u_{\gamma'}}{\partial K}}{u_\gamma} +  O(n^{-1}) \right\}.
\label{Tr_k}
\end{eqnarray}
The quantity $O(n^{-1})/n$ in Eq. (\ref{Tr_k}) when summed over all values of 
occupied $k$'s (they are $O(n)$) gives a contribution $O(n^{-1})$ vanishing for 
$n \rightarrow \infty$.
Therefore, if no partially filled bands are present, one derives the following
formula for each spin:
\begin{eqnarray}
\lim_{n \rightarrow \infty} \frac{Tr(\mathbf{\Lambda})_\sigma}{n} & = &
\frac{d}{2 \pi }  \int_{K_1}^{K_2}  \left( 
\sum_{\gamma}
\scal{\frac{\partial u_{\gamma}}{\partial K}}{\frac{\partial u_{\gamma}}{\partial K}}+ \right. \nonumber \\
&& \!\!\!\!- \left.\sum_{\gamma \gamma'}  \!\!
\scal{u_\gamma}{\frac{\partial u_{\gamma'}}{\partial K}} \!\!
\scal{\frac{\partial u_{\gamma'}}{\partial K}}{u_\gamma} \!\! \right)\!\!  dK,
\label{sgperefla}
\end{eqnarray} 
where we replaced $\sum_k $ by $\frac{nd}{2 \pi} \int_{K_1}^{K_2} \ldots dK$. 
In case of $n_b$ doubly
occupied bands we have $2 n_b$ electrons per cell, but Eq. (\ref{sgperefla}) should be multiplied by
2 to account for both spins. The final result is Eq. (\ref{sgperefla}) divided by $n_b$ which is 
nothing but Eq. (16) of the paper  by  C. Sgiarovello {\it et al.}.\cite{sgpere} The latter was 
obtained  by working out the formalism of Resta {\it et al.} \cite{RestaSorella} for a 
determinantal wavefunction with PBC. 

\subsubsection{Sum over states.}
\label{sec:sos}

By inserting a completeness of the virtual space in Eq. (\ref{trace}),  it can be rewritten as follows:
\begin{equation}
Tr(\mathbf{\Lambda}) \: = \:
\sum_{jl}  \sum_{m  v} \: \scal{\Phi_{j}^{\tilde{\jmath}}}{\Phi_{m}^v}                        \scal{\Phi_{m}^v}{\Phi_{l}^{\tilde{l}}},
\label{sumovst}
\end{equation}
where multi indexes $j, l, m$ run 
over occupied spin orbitals and $v$ over virtual 
ones, $v \, = \, \{k_v,\eta,\sigma\}$.
Given that $\scal{\Phi_m^v}{\Phi_{l}^{\tilde{l}}}=   \scal{\psi_v}{\tilde{\psi}_l} \delta_{lm}$
and using Eq. (\ref{matrel}),  we realize that the previous expression contains the
factor $ \delta_{kk'}\delta_{\gamma \gamma'} \delta_{k_v,k \mp 1}$.
In this way we get, for each spin $\alpha$ and $\beta$, the following contribution to 
$Tr(\mathbf{\Lambda})$:
\begin{eqnarray}
Tr(\mathbf{\Lambda})_\sigma \!& = & \!\!
\sum_{k, \gamma, \eta }  \!\! 
\scal{ (x \pm iy) \psi_{k \gamma}}{\psi_{k \pm 1, \eta}} 
\scal{\psi_{k \pm 1, \eta }}{(x \pm iy) \psi_{k \gamma}} \nonumber \\
 & = &  
\left( \frac{nd}{2 \pi}\right)^2 \sum_{k \gamma \eta}
  \left| \int_{0}^d  [ u_{k ,\gamma}(s) ]^* u_{k \pm 1,\eta}(s)  ds \right| ^2 \nonumber \\
 & = &  
 \sum_{k \gamma} \, \left[  \sum_{\eta} \left( \frac{nd}{2 \pi}\right)^2
  \left| \langle u_{k ,\gamma} |  u_{k \pm 1,\eta} \rangle \right| ^2 \right]\nonumber \\
   & = & \sum_{\gamma k} \; Tr(\mathbf{\Lambda})_{\gamma k \sigma},
\label{sumoverstates}
\end{eqnarray}
where $k, \, \gamma$ run over occupied spin orbitals of the given spin $\sigma$
and $\eta$ over virtual ones. Eq. (\ref{sumoverstates}) gives an alternative
expression of the contribution of each spin orbital and can be used to
numerically compute $Tr(\mathbf{\Lambda})$ for a given value of $n$. It should also
be reminded that the sum over virtual orbitals is in principle infinite, because
the expansion of $\Phi_{j}^{\tilde{\jmath}}$ in single excitations is exact
in general only when the orbital basis is complete. This condition is not
in general fulfilled in actual calculations of LCAO type and this amounts to an
approximation. An exception is the H\"uckel method where Eqs. (\ref{sumoverstates})
and (\ref{preresta}) or (\ref{direct}) are strictly equivalent as a 
result of particular assumptions about the orbital basis.
In order to examine Eq. (\ref{sumoverstates}), we refer to Eq. (\ref{intsquare})
 and point out the presence of  $\delta_{\gamma \eta}$ in the right-hand side. Suppose there is
a band $\gamma$ not completely filled: virtual band index $\eta$ can assume 
the value $\gamma$ and generates a diverging contribution $(nd/2 \pi)^2+O(1)$ for
$n \rightarrow \infty$.  In this way, we show again the
equivalence of two criteria for establishing the metallic-insulating 
character of a system, namely: 1) fractionally filled band 
2) divergence for $n \rightarrow \infty$ of the TPS/number of electrons. 
\\

Let us now consider a system with completely filled bands (insulator), for which
$\delta_{\gamma \eta}=0$ always. We replace the $\sum_{k}$ by 
$\frac{nd}{2 \pi} \: \int_{K_1}^{K_2} \: dK$ with $K_2-K_1= 2 \pi/d$ and obtain the final 
result for the contribution of each spin to $Tr{(\mathbf{\Lambda})}$:
\begin{equation}
Tr{(\mathbf{\Lambda})_\sigma} \;  = \; 
 \frac{nd}{2 \pi} \: \sum_{\gamma \eta} \:\int_{K_1}^{K_2}
\scal{u_{\gamma}}{\frac{\partial u_{\eta}}{\partial K}} 
\scal{\frac{\partial u_{\eta}}{\partial K}}{u_{\gamma}} \: dK. 
\label{sosfinal}
\end{equation}
It is clear from Eq. (\ref{sosfinal}) that the TPS diverges for $n \rightarrow \infty$, as 
expected. 
The TPS per electron is obtained by dividing Eq. (\ref{sosfinal}) by the number
of electrons $n_e$. The latter is proportional to $n$; it can be expressed as a 
function of the density $\rho$ as $n_e=nd\rho=L\rho$ or of the number of occupied 
bands $n_b$ times their occupation number $n_o$ (1 or 2) and the number of addends 
$n_k$ in the $\sum_k$.
For a system with only doubly filled bands one has:
 $n_k=n, \; n_o=2$, $n_e=2 \, n \, n_b$ 
\begin{eqnarray}
&&\!\!\!\frac{Tr{(\mathbf{\Lambda})_\alpha}+Tr{(\mathbf{\Lambda})_\beta} }{2 \, n \, n_b} =  \nonumber \\
 &&\quad\quad=
\!\frac{d}{2 \pi n_b}\!  \sum_{\gamma \eta} \!\int_{- \pi/d}^{ \pi/d}
\!\scal{\!u_{\gamma}}{\frac{\partial u_{\eta}}{\partial K}} 
\!\scal{\!\frac{\partial u_{\eta}}{\partial K}}{u_{\gamma}}  dK ,
\label{sosiso}
\end{eqnarray} 
where $n_b$ is the number of doubly occupied bands, $\gamma$ runs over occupied and
$\eta$ over virtual bands.

\subsection{The polarizability.}
\label{sec:polarizability}

The static dipole polarizability tensor is given by:\cite{Hirschfelder}
\begin{equation}
 {\mathbf \alpha}_{xy} \; = \; 
2 \elm{\Phi_0}{\mu_x\left( {\mathbf H} - E_0 \right)_{\perp}^{-1}\mu_y}{\Phi_0},
\label{polarizability}
\end{equation}
where $\left( {\mathbf H} - E_0 \right)_{\perp}^{-1}$ is the reduced resolvent
of the Hamiltonian in the orthogonal complement to $\Phi_0$.

Let us consider the quantity:
\begin{eqnarray}
 \mathbf{\alpha} & = &
 \elm{(\hat{X} \pm i \hat{Y}) \Phi}{(\mathbf{H}_K-E_0)_{\perp}^{-1}}{(\hat{X} \pm i \hat{Y}) \Phi}  \: = \nonumber \\
   & = & \elm{\hat{X}\Phi}{(\mathbf{H}_K-E_0)_{\perp}^{-1}}{\hat{X} \Phi}+ \nonumber \\
   & & +
              \elm{\hat{Y}\Phi}{(\mathbf{H}_K-E_0)_{\perp}^{-1}}{ \hat{Y} \Phi} + \nonumber \\
 &  & \pm i \elm{\hat{X}\Phi}{(\mathbf{H}_K-E_0)_{\perp}^{-1}}{ \hat{Y} \Phi} + \nonumber \\
 & &   \mp i \elm{\hat{Y}\Phi}{(\mathbf{H}_K-E_0)_{\perp}^{-1}}{ \hat{X} \Phi}.
\label{eq:XYpolariz}
\end{eqnarray}
As shown in the Appendix, the first two terms of Eq. (\ref{eq:XYpolariz}) are
equal, while the last two are vanishing; this allows us to write:   
\begin{equation} 
{\mathbf \alpha}_{xx} \; = \;  \elm{\sum_{j} \Phi_{j}^{\tilde{\jmath}}}{(\mathbf{H}_K-E_0)_{\perp}^{-1}}{\sum_{j'} \Phi_{j'}^{\tilde{\jmath}'}}.
\label{alphaxx}
\end{equation}
In the subspace of single excitations one has:
\begin{equation}
(\mathbf{H}_K-E_0)^{-1} \; = \; \sum_{j,v} \, 
\frac{ | \Phi_{j}^v \rangle \langle \Phi_{j}^v | }{ \epsilon_v - \epsilon_j} \, ; \\
\end{equation}
and, for finite $n$ and a  given spin $\sigma$:
\begin{eqnarray}
\mathbf{\alpha}_{xx \sigma} & = & \sum_{k, \gamma, \eta } \;  
\frac{\scal{(x \pm iy) \psi_{k \gamma}}{\psi_{k \pm 1, \eta}} 
\scal{\psi_{k \pm 1, \eta }}{(x \pm iy) \psi_{k \gamma}}}
{ \epsilon_{k \pm 1,\eta} - \epsilon_{k,\gamma}} \nonumber \\ 
\quad & = &  
\left( \frac{nd}{2 \pi}\right)^2 \sum_{k \gamma \eta}
  \frac{ \left| \int_{0}^d \: [ u_{k ,\gamma}(s) ]^* u_{k \pm 1,\eta}(s) \: ds \right| ^2}{ \epsilon_{k \pm 1,\eta} - \epsilon_{k,\gamma} } \nonumber \\ 
\quad & = &  
\left( \frac{nd}{2 \pi}\right)^2 \sum_{k \gamma \eta}
  \frac{\left| \scal{u_{k ,\gamma}}{u_{k \pm 1,\eta}} \right|^2}{ \epsilon_{k \pm 1,\eta} - \epsilon_{k,\gamma} }.
\label{sospolarizability}
\end{eqnarray}
For large $n$ we switch to the $K$ variable also for $\epsilon$ ($\epsilon_{k,\eta} \; \leftrightarrow \; \epsilon_{\eta}(K)$, see Eq. \ref{kcontinuo}). From Eq. (\ref{intsquare}) and
provided that $\epsilon_{k,\eta} \ne \epsilon_{k,\gamma}$ one has:
\begin{eqnarray}
&&(\epsilon_{k+1,\eta}-\epsilon_{k,\gamma})^{-1} = 
(\epsilon_{\eta}(K)-\epsilon_{\gamma}(K))^{-1} +\nonumber \\
&&\quad\quad\quad\quad -
\frac{2 \pi}{nd} (\epsilon_{\eta}(K)-\epsilon_{\gamma}(K))^{-2}
\frac{\partial \epsilon_{\eta}}{\partial K}  + O(L^{-2}) 
\end{eqnarray}
and we get: 
\begin{eqnarray}
&&\frac{\left| \scal{u_{k ,\gamma}}{u_{k \pm 1,\eta}}\right| ^2}{ \epsilon_{k \pm 1,\eta} - \epsilon_{k,\gamma} }   = 
\frac{\delta_{\gamma \eta}}{\epsilon_\eta(K)-\epsilon_\gamma(K)} \;  + 
\nonumber \\
& &\quad\quad\quad + 
\frac{2 \pi}{nd} 
\frac{\scal{u_\gamma}{\frac{\partial u_\eta}{\partial K}} 
\scal{\frac{\partial u_\eta}{\partial K}}{u_\gamma} }{\epsilon_\eta(K)-\epsilon_\gamma(K)} +
\nonumber \\
& &\quad\quad\quad + 
\delta_{\gamma \eta} \frac{2 \pi}{nd} 
\frac{\partial \epsilon_\eta}{\partial K} \, \frac{1}{(\epsilon_\eta(K)-\epsilon_\gamma(K))^2} 
+ \nonumber \\
&  & \quad\quad\quad+
	\left( \frac{2 \pi}{nd} \right)^2 \, \frac{\partial \epsilon_\eta}{\partial K} \, 
\frac{ \scal{u_\gamma}{\frac{\partial u_\eta}{\partial K}}
\scal{\frac{\partial u_\eta}{\partial K}}{u_\gamma}}{ (\epsilon_\eta(K)-\epsilon_\gamma(K))^2}  + \nonumber \\
 & & \quad\quad\quad-\delta_{\gamma \eta}
\left( \frac{2 \pi}{nd}  \right)^2 \!\! 
\frac{\scal{\frac{\partial u_\gamma}{\partial K}}
{\frac{\partial u_\gamma}{\partial K}}}
{\epsilon_\eta(K)-\epsilon_\gamma(K) } 
+ O(L^{-3}).
\end{eqnarray} 
\\

In the case of a band insulator with a gap separating the occupied band $\gamma$ 
from the virtual one $\eta$, the polarizability per unit cell is given by:
\begin{equation}
\alpha_{xx} \; = \; \frac{d}{\pi}  
 \:  \sum_{\gamma \eta} \:\int_{- \pi/d}^{ \pi/d}
	\frac{\scal{u_{\gamma}}{\frac{\partial u_{\eta}}{\partial K}}
	      \scal{\frac{\partial u_{\eta}}{\partial K}}{u_{\gamma}}}
	      {\epsilon_\eta(K)-\epsilon_\gamma(K)} \: dK .
	\label{polarizability2} 
\end{equation}
In the case of  a partially filled band $\gamma=\eta$, the denominator 
vanishes at $k=k_F$ and the polarizability diverges. 

\section{Examples}
\label{examples}

Formulas (\ref{sgperefla}) and (\ref{sosfinal},) can be used for numerical computation
in general but in case of exactly solvable models, a symbolic evaluation 
is possible. 
Here we consider  two examples: the H\"uckel model of dimerized annulene and 
that of cyclacene.
The orbitals $\psi$ are linear combinations of site functions $\chi(P)$ 
centered in point $P$ as previously pointed out in Sec. \ref{approxwf}.

\subsection{Dimerized annulene.}
\begin{figure}[htb]
\centering
\includegraphics[scale=1]{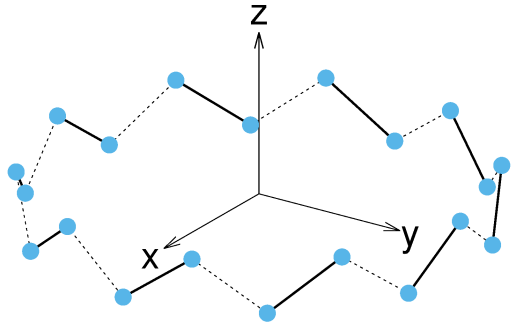}
\caption{Geometry of dimerized annulene for $n=10$. Full lines and dashed lines connecting the C atoms indicate the two resonance integrals,
$\beta_1$ and $\beta_2$, respectively.}
\label{fig:dim-ann}
\end{figure}

The H\"uckel model for dimerized annulene of length $L=nd$ consists of $n$ 
units or cells, each containing two sites and one electron per site.  
The sites are assumed to be equally separated but connected by bonds of 
different strength described by two resonance integrals $\beta_1, \;\beta_2$.
The dimerization is parametrized by $\delta$ in such a way that the non 
dimerized case is recovered at $\delta=0$, as detailed in the following. 
A schematic representation of a dimerized annulene with $n=10$ is reported in Fig. \ref{fig:dim-ann}.

The orbitals are given by:
\begin{equation}
\psi_k(x,y) =  \frac{1}{\sqrt{n}} 
\sum_{\mu=0}^{n-1} e^{\frac{ 2 \pi i k \mu}{n}}
\left[ c_1 \, \chi(P_{1\mu})  + c_2 \, \chi(P_{2\mu}) \right],
\end{equation}     
where $\mu$ is the cell index and the coordinates of the centers are:  
\begin{equation*}
\begin{array}{l|cc}
\quad   &     x      &     y   \\
\hline
\quad & \quad & \quad \\
P_{1\mu} &   R \cos \frac{\mu d}{R}         &  R \sin \frac{\mu d}{R} \\
\quad & \quad & \quad \\
P_{2\mu} &   R \cos \frac{(\mu  + 1/2)d}{R}    &  R \sin \frac{(\mu + 1/2)d }{R} 
\end{array}
\label{Pcoord}
\end{equation*} 
\\ 

The coefficients $c_1$ and $c_2$ are  obtained by diagonalizing the Hamiltonian matrix 
$\mathbf{H}_k$ defined in Eq. (\ref{hamvar}) and reported in Table \ref{tab:hampoly},
\begin{table}
\caption{Effective Hamiltonian matrix for dimerized annulene.}
\[
\mathbf{H}_k \; = \; \left[
\begin{array}{cc}
\alpha  &  \beta_1 e^ {\frac{- 2 \pi i k}{n}} + \beta_2\\
\quad & \quad \\
\beta_1 e^{ \frac{ 2 \pi i k}{n}}+ \beta_2 & \alpha   
\end{array} \right]
\]
\label{tab:hampoly}
\end{table}
where $\beta_1 \, = \, -t(1+\delta),\beta_2 \, = \, -t(1-\delta)$ and $t>0$ is 
the hopping integral of the undimerized annulene. 
In Table  \ref{tab:eigensystem} we report eigenvalues $\epsilon$ and eigenvector
components $\{ c_1, \, c_2 \}$ of the 
matrix reported in Table \ref{tab:hampoly} for $\alpha=0$.
\begin{table}
\caption{Eigenvalues and normalized eigenvectors of the matrix $\mathbf{H}_k$ of
	dimerized annulene.}
\[
\begin{array}{l|c|c}
\epsilon & -t \sqrt{2[1+\delta^2+(1-\delta^2) \cos \varkappa ]}& 
            t \sqrt{2[1+\delta^2+(1-\delta^2) \cos \varkappa ]} \\
\hline 
\quad & \quad & \quad \\
c_1   & 
\frac{e^{i \varkappa}(\delta-1)-\delta-1}{\sqrt{2}(e^{i \varkappa}(\delta+1) - \delta + 1} & 
- \frac{e^{i \varkappa}(\delta-1)-\delta-1}{\sqrt{2}(e^{i \varkappa}(\delta+1) - \delta + 1}\\  
\quad & \quad & \quad \\
c_2 & 
\frac{e^{i \varkappa}(\delta-1)-\delta-1}{2 \sqrt{1+\delta^2+(1-\delta^2) \cos \varkappa } } &
\frac{e^{i \varkappa}(\delta-1)-\delta-1}{2 \sqrt{1+\delta^2+(1-\delta^2) \cos \varkappa }} \\
\end{array}
\]
\label{tab:eigensystem}
\end{table}
\\

We used the variable $\varkappa \; = \; 2 \pi k / n$ related to $K$ by 
$K=\varkappa / d$. According to Eqs. (\ref{ukpiccolo}) and (\ref{ukgrande}) the periodic 
part of the H\"uckel orbital is given in cell $\mu$ by:
\begin{eqnarray}
u_k(s) & = & c_1 \chi(P_{1\mu})  +  e^{-i \pi k/n}  c_2  \chi(P_{2\mu}),  \\ 
u(s,K) & = & c_1 \chi(P_{1\mu})  +  e^{-i K d/2}  c_2  \chi(P_{2\mu}), 
\end{eqnarray}
where it should be reminded that $c_1$ and $\,c_2$ are functions of $k$ or $K$.
\\

\begin{figure}[htb]
\centering
\includegraphics[scale=.65]{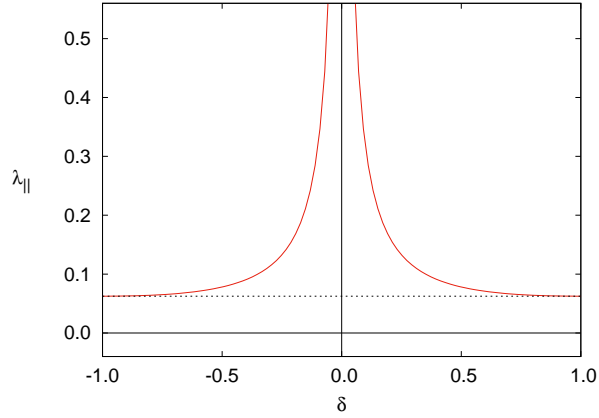}
\caption{TPS per electron of dimerized annulene as a function of $\delta$.}
\label{fig:polydim}
\end{figure}

Eqs. (\ref{sgperefla}) and (\ref{sosiso}) were both symbolically computed using 
MATHEMATICA 12.1 \cite{mathematica} and gave identical results: 
\begin{equation}
\frac{Tr(\mathbf{\Lambda})_\alpha + Tr(\mathbf{\Lambda})_\beta}{ 2 n } \; = \;
\frac{ d^2 (1+ \delta^2)}{32 |\delta|} .
\label{deltadim}
\end{equation}
This result has been reported also in Ref. [\onlinecite{Valenca}], where a factor 16 at the denominator is reported  instead of 32, therefore the TPS per unit is given there instead of the TPS per electron.
Eq. (\ref{deltadim}) is reported in Fig. \ref{fig:polydim} for $d=1$.


The limit $\delta \rightarrow 0$ is $+\infty$ as expected for a conductor, while for $\delta = \pm 1$
one gets $d^2/16$ which is the value of a molecule composed of two sites at the distance $d/2$.
The TPS of such a system with one electron sitting on each site is $(d/4)^2 + (-d/4)^2$ to be
divided by 2 electrons.
\\

\begin{figure}[htb]
\centering
\includegraphics[scale=.65]{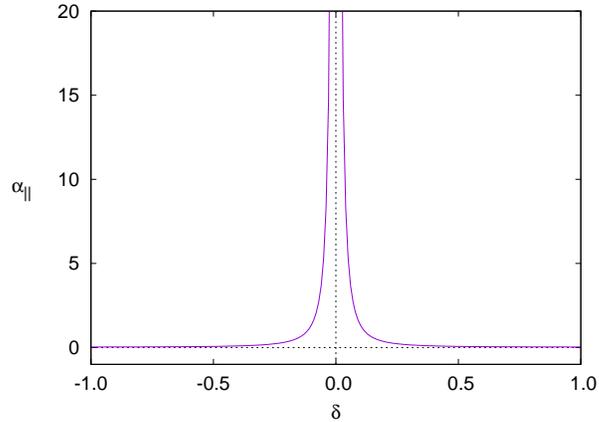}
\caption{$\alpha_\parallel$ per cell of dimerized annulene as a function of $\delta$.}
\label{fig:alpha_polyac}
\end{figure}

As concerns the polarizability we find:
\begin{equation}
	\alpha_{\parallel} \; = \; \frac{ 2(1+\delta^2)E(1-\delta^2)-\delta^2 K(1-\delta^2)}{48 \pi \delta^2} \, \frac{d^2}{t}, 
\end{equation}
where $K$ and $E$ are the complete elliptic integrals of the first and second kind,
respectively:
\begin{equation}
K(x) \; = \; \int_{0}^{\pi/2} \left( 1 - x \, \sin^2 \theta \right)^{-1/2}  d\theta 
\label{eq:k}
\end{equation}
and
\begin{equation}
E(x) \; = \; \int_{0}^{\pi/2} \left( 1 - x \, \sin^2 \theta \right)^{1/2}  d\theta .
\label{eq:e}
\end{equation} 
In Fig. \ref{fig:alpha_polyac} we report $\alpha_\parallel$ as a function of 
$\delta$ for $d=t=1$.

\subsection{Cyclacene} 
\begin{figure}[htb]
\centering
\includegraphics[scale=.65]{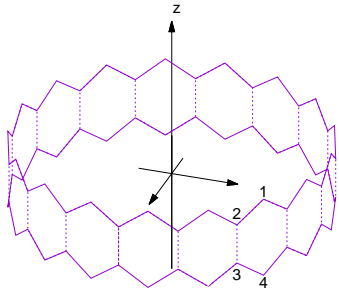}
\caption{Geometry of cyclacene for $n=17$. Full lines and dashed lines connecting the C atoms indicate the two different hopping integrals, $t$ and $\eta t$, respectively.}
\label{fig:cyclacene}
\end{figure}

The geometry of cyclacene is assumed to be a strip of $n$ regular hexagons 
folded in a cylinder see Fig. \ref{fig:cyclacene}; the axis of the cyclacene ring  
is $z$.
The length of the elementary cell is $d=b \sqrt{3}$ where $b$ is the side of the
hexagon. The coordinates of the sites are given in Table \ref{Ccoord}.
\begin{table}
\caption{Coordinates of the sites ($\mu=0,\cdots,n-1$) for the cyclacene molecule.\label{Ccoord}}
\begin{tabular}{l|ccc}
	\quad   &     x      &     y   &  z  \\
\hline
\quad & \quad & \quad & \quad \\
	$P_{1\mu}$ &   $R \cos \frac{(\mu  + 1/2)d}{R}$ &  $R \sin \frac{(\mu + 1/2)d }{R}$ & $b$ \\
\quad & \quad & \quad & \quad \\
	$P_{2\mu}$ &   $R \cos \frac{\mu d}{R}$         &  $R \sin \frac{\mu d}{R}$ & $\frac{b}{2}$ \\
\quad & \quad & \quad & \quad \\
	$P_{3\mu}$ &  $ R \cos \frac{\mu d}{R}$         &  $R \sin \frac{\mu d}{R}$ & $- \frac{b}{2}$ \\
\quad & \quad & \quad & \quad \\
	$P_{4\mu}$ & $  R \cos \frac{(\mu  + 1/2)d}{R}$ &  $R \sin \frac{(\mu + 1/2)d }{R}$  & $-b$ 
\end{tabular}
\end{table} 
The cyclacene molecule  is symmetric with respect to the $x,y$ plane and can be viewed as
two annulene rings, one above and one below this $\sigma_h$ plane, connected by bonds 
parallel to the $z-$axis, as shown in Fig. \ref{fig:cyclacene} by dashed lines. 
The effective Hamiltonian matrix is given in Table \ref{hamcyclacene}, where we 
considered the possibility of a different strength for the vertical bonds connecting
the two annulene rings by introducing a parameter $0 \le \eta \le 1$. The value $\eta=1$ 
corresponds to the cyclacene molecule while for $\eta=0$ one gets two non interacting 
and undimerized annulenes. 
\begin{table}
\caption{Effective Hamiltonian matrix for cyclacene; $t>0$ is the hopping integral.}
\[
	\left[ \begin{array}{cccc}
\alpha & \!\!\!\!-t\!\left(1\!+\!e^{\frac{2 \pi i k}{n}} \right)\!\!\!\! & 0 & 0 \\
\quad & \quad & \quad & \quad \\
	\!\!\!\!-t\!\left(1\!+\!e^{\frac{ -2 \pi i k}{n}} \right)\!\!\!\! & \alpha  & -\eta t & 0  \\
\quad & \quad & \quad & \quad \\
	0 & -\eta t & \alpha & \!\!\!\!-t\!\left(1+e^{\frac{ -2 \pi i k}{n}} \right)\!\!\!\!   \\
\quad & \quad & \quad & \quad \\
0 & 0 & \!\!\!\!-t\!\left(1+e^{\frac{ 2 \pi i k}{n}} \right)\!\!\!\! & \alpha   
\end{array} \right] 
\]
\label{hamcyclacene}
\end{table}
The eigenvalues are reported in Table \ref{eigcyclacene}; the eigenvectors are not reported because they are exceedingly complicated, but they can be found in Appendix II.
\begin{table}
\caption{Eigenvalues of cyclacene. $\Upsilon=8+\eta^2+8\cos \frac{ 2 \pi k}{n}$.}
\[ \begin{array}{l|c|c|c|c|}
\quad & \epsilon_1 & \epsilon_2 & \epsilon_3 & \epsilon_4 \\
\hline 
\quad & \quad & \quad & \quad & \quad \\ 
 \sigma_h & + & - & + & - \\
\quad & \quad & \quad & \quad & \quad \\ 
\mbox{energies} & -t\frac{\eta +\sqrt{\Upsilon}}{2} & t\frac{\eta -\sqrt{\Upsilon}}{2} & t\frac{\sqrt{\Upsilon}-\eta}{2} & t\frac{\sqrt{\Upsilon}+\eta}{2} 
\end{array} \]
\label{eigcyclacene}
\end{table}
\\

\begin{figure}[htb]
\centering
\includegraphics[scale=.65]{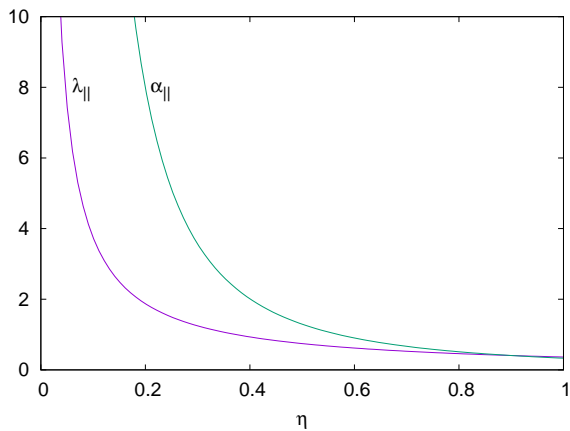}
\caption{TPS per electron and polarizability of parametrized cyclacene as a function of $\eta$. The units
are $b^2$ and $b^2/t$ for $\lambda_\parallel$ and $\alpha_\parallel$, respectively.}
\label{fig:lampolcycl}
\end{figure}

In Eq. (\ref{eq:cyclaresults}) we report the localization spread and polarizability
per cell of cyclacene. 
The TPS per electron was computed using Eqs. (\ref{sgperefla}) or (\ref{sosiso})  and 
the polarizability per cell using Eq. (\ref{polarizability}), obtaining:
\begin{eqnarray}
\lambda_\parallel(\eta) & = & \frac{ 3 }{2 \eta \sqrt{16+\eta^2}} b^2,  \nonumber \\
	\alpha_\parallel(\eta)  & = & \frac{1}{ 8 \pi \sqrt{16+\eta^2}} \left[ \frac{ 32+\eta^2}{\eta^2} E\left( \frac{16}{16+\eta^2} \right) + \right. 
	\nonumber \\
	& & \left. - K \left( \frac{16}{16+\eta^2} \right) \right] \, \frac{b^2}{t}, 
\label{eq:cyclaresults}
\end{eqnarray}
where $K(x)$ and $E(x)$ are defined in Eqs. (\ref{eq:k}) and (\ref{eq:e}).  In Fig. \ref{fig:lampolcycl} we report the results given in Eq.
(\ref{eq:cyclaresults}). 
\\

Both $\lambda_\parallel$ and $\alpha_\parallel$ diverge for $\eta \rightarrow 0$ as expected for
a couple of metallic annulenes. On the other hand 
at $\eta=1$ we obtain the following results for the cyclacene molecule:
\[ 
\begin{array}{l}
\lambda_\parallel(1) \; = \; \frac{ 3 }{2  \sqrt{17}} b^2 \; \approx \; 0.363804  \,  b^2,\\
\quad  \quad \\
	\alpha_\parallel(1)   \; = \; \frac{ 33 E \left( \frac{16}{17} \right)  - K \left( \frac{16}{17} \right) }{8 \pi \sqrt{17}} \, \frac{b^2}{t} \; \approx \; 0.313082 \, \frac{b^2}{t} ,  
\end{array}
\]
showing its insulating character in the $xy$ plane and recovering the results found 
in a previous paper.\cite{cyclacene}

\section{Discussion and conclusions.}

In this paper we exploit the isomorphism between the $C_n$ group and the group of 1D
translations with periodic Born von K\'arm\'an boundary conditions. We consider a finite 
ring of radius $R$ with open boundary conditions in the $(x,y)$ plane and a segment of
length $L=2 \pi R$  on a straight line with periodic boundary conditions. If we denote by $\phi$
the rotation angle around the centre of the ring in the \textit{e.g.} counterclockwise direction, 
the arc length $s=R \phi$ on the ring is mapped on the coordinate, say  $\zeta$, on the line segment 
counted from \textit{e.g.} its leftmost point: $ 0 \le \zeta < L$. 
This can be viewed as rolling the ring on the straight line and in this sense all the
points of the line can be mapped on the ring provided the angle $\phi$ is allowed to assume
any real value. 
In the plane we can use a single complex coordinate $z=x+iy$ to describe any curve and in this
way the equation of the ring is $z=R (\cos \phi + i \sin \phi) = R \exp (i \phi)= R \exp ( i s / R)$. 
The point $P(z)$ of the ring is mapped on the point $P(\zeta)$ on the line, and we recover 
the {\it complex position} operator introduced in Ref. [\onlinecite{Valenca}]. 
This mapping provides new insight into the nature of the complex position operator.
\\

As far as the TPS is concerned,  we can easily derive formulae for the thermodynamic limit for systems 
treated at non-correlated level, \textit{i.e.} described by a Slater determinant. In particular  a 
formula of Resta and coworkers is obtained in a different way from the original derivation.\cite{sgpere}
More interesting a second formula, we called {\it sum-over-states}, for the TPS, equivalent to the Resta 
one in the limit of a complete basis, is also obtained. The latter allows for an interesting extension to 
the polarizability and to any quantity expressed as: 
\begin{equation}
\langle \Phi_0 \mu_x \left( {\mathbf H} - E_0 \right)_{\perp}^{K+1} \mu_x \Phi_0 \rangle =  S_K .
\end{equation}
The quantities $S_K$ have been the object of much interest in the old days of perturbation theory 
\cite{Hirschfelder} and are known as sum rules for oscillator strength.
As already pointed out in Ref. [\onlinecite{Valenca}] our approach using the complex position one-body operator
can be applied to metallic systems avoiding the awkward ``$\ln 0$'' singularity. This allows us to compute 
$\lambda$ for finite systems and study their behaviour when approaching the thermodynamic limit. 
As discussed in sections \ref{sec:direct} and \ref{sec:sos} the divergence of $\lambda$ is due to the partial 
filling of a band: this shows the equivalence of the two criteria for a non correlated system to be a conductor.
\\

Finally, we want to stress the fact that, our approach is not confined to the treatment of periodic non interacting systems, although this was the subject of the present work.
Indeed, once the ordinary position operator is replaced by the periodic complex-position one, it is possible to proceed exactly as in the case of OBC. It is worth noticing that the use of the periodic complex-position operator does not introduce complications for the numerical evaluation of its mean value, given that it is the square of a one-electron operator, exactly as in the case of the ordinary position operator.
In all the cases we have investigated so far, the large-system qualitative behavior of the real and complex-position quantities is identical.
\\

Concerning the treatment of correlated systems, we note that our approach does not present peculiar problems, given that one has to evaluate the mean value of the square of a one-electron operator and the machinery of quantum chemistry can be easily adapted to perform this task (playing attention to the fact that the operator is in this case complex). Actually, we have already treated correlated systems following the approach here reported.\cite{Valenca}
The difficulty, which is general for any approach,  is mainly a ``technical'' one, since it is very hard to compute correlated wave-functions for systems having more than a dozen identical units.
In a similar way, it will be possible to treat disordered systems, exactly in the same way done by using finite OBC formalism.\cite{Benda10}

Finally, we notice that the extension of the formalism to 2D and 3D systems will be the subject of future work.


\section*{Appendix 1}
In this Appendix we show the vanishing  of some matrix
elements for systems enjoying the symmetry of the $C_n$ group.
In particular we consider the matrix elements of Eq. (\ref{eq:XYpolariz})
and use group theory arguments.
Let us consider first the functions defined by the cartesian coordinates
$x,y$ of a point $P$ in the ring, see the Supplementary material. 
We rewrite the latter as:
\begin{eqnarray}
  x(P) & = & \frac{R}{2}  \left(e^{ \frac{2 \pi i s}{L}} +
              e^{ \frac{- 2 \pi i s}{L}}\right),  \\
  y(P) & = & -i \frac{R}{2} \left( e^{ \frac{2 \pi i s}{L}} -
              e^{ \frac{- 2 \pi i s}{L}}\right),  
\label{eq:xandy}
\end{eqnarray}
and, by comparison with Eq. (\ref{bloch}), we realize that this couple
of functions belong to the reducible  representation E with
$k=\pm 1$. Therefore,  expectation values of the dipole operators in
the ring wavefunctions are vanishing.
As concerns the second moments, we have: 
\begin{eqnarray}
  x^2(P) & = & \frac{R^2}{4} \left[ e^{\frac{4 \pi i s}{L}} +
e^{\frac{- 4 \pi i s}{L}} + 2 \right], \\
  y^2(P) & = & - \frac{R^2}{4} \left[ e^{\frac{4 \pi i s}{L}} +
e^{\frac{- 4 \pi i s}{L}} - 2 \right], \\
xy(P) & = & -i \frac{R^2}{4} \left[ e^{\frac{4 \pi i s}{L}} +
 e^{\frac{- 4 \pi i s}{L}} \right].
\end{eqnarray}
Therefore $xy$ and $x^2-y^2$ belong to the reducible representation $k=\pm 2$, 
while $x^2$ and $y^2$ contain the A representation ($x^2+y^2$ 
belong to A).

\section*{Appendix II}

Here we report the eigenvalues and eigenvecors of the system of two annulenes
coupled to form a cyclacene molecule when the parameter $\eta$ is equal to $1$


\begin{table*}
\caption{Eigenvalues and normalized eigenvectors of two weakly bonded annulenes forming cyclacene for $\eta=1$.
  $\Upsilon=8+\eta^2+8 \cos \frac{ 2 \pi k}{n}$.}
\[ \begin{array}{l|c|c|}
\quad & \quad & \quad \\ 
\quad & \mbox{eigenvector} \; 1 & \mbox{eigenvector} \; 2 \\ 
\quad & \quad & \quad \\ 
\mbox{energy} & -t(\eta+\sqrt{\Upsilon})/2 & t(\eta-\sqrt{\Upsilon})/2 \\ 
\quad & \quad & \quad \\ 
 \sigma_h & + & - \\ 
\hline 
\quad & \quad & \quad \\ 
c_1 & \frac{( \Upsilon - \eta \sqrt{\Upsilon}) \sqrt{ \Upsilon+\eta \sqrt{\Upsilon}} }{ 8 \Upsilon \cos (k \pi/n) } & 
 -  \frac{( \Upsilon + \eta \sqrt{\Upsilon}) \sqrt{ \Upsilon-\eta \sqrt{\Upsilon}} }{ 8 \Upsilon \cos (k \pi/n) } \\
\quad & \quad & \quad \\ 
c_2 & \frac{ \sqrt{ \Upsilon+\eta \sqrt{\Upsilon}} \sqrt{\Upsilon^2-\eta^2 \Upsilon} ( 1 - i \tan (k \pi / n) ) }{ 32 \Upsilon \cos (k \pi /n ) } &  
    \frac{ \sqrt{-\Upsilon+\eta \sqrt{\Upsilon}} \sqrt{\Upsilon^2-\eta^2 \Upsilon} ( 1 + i \tan (k \pi / n) ) }{ 32 \Upsilon \cos (k \pi /n ) } \\  
\quad & \quad & \quad \\ 
c_3 & \frac{ \sqrt{ \Upsilon+\eta \sqrt{\Upsilon}} \sqrt{\Upsilon^2-\eta^2 \Upsilon} ( 1 - i \tan (k \pi / n) ) }{ 32 \Upsilon \cos (k \pi /n ) } &
    \frac{( \Upsilon + \eta \sqrt{\Upsilon}) \sqrt{ \Upsilon-\eta \sqrt{\Upsilon}} }{ 8 \Upsilon \cos (k \pi/n) } \\
\quad & \quad & \quad \\
c_4 & \frac{( \Upsilon - \eta \sqrt{\Upsilon}) \sqrt{ \Upsilon+\eta \sqrt{\Upsilon}} }{ 8 \Upsilon \cos (k \pi/n) } &
    \frac{( \Upsilon + \eta \sqrt{\Upsilon}) \sqrt{ \Upsilon-\eta \sqrt{\Upsilon}} }{ 8 \Upsilon \cos (k \pi/n) } \\
\quad & \quad & \quad \\ 
\hline
\hline
\quad & \quad & \quad \\
\quad & \mbox{eigenvector} \; 3 & \mbox{eigenvector} \; 4 \\ 
\quad & \quad & \quad \\ 
\mbox{energy}  & t(\sqrt{\Upsilon}-\eta)/2 & t(\sqrt{\Upsilon}+\eta)/2 \\
\quad & \quad & \quad \\
\sigma_h & + & - \\
\hline
	\quad & \quad & \quad \\
c_1 &  \frac{( \Upsilon + \eta \sqrt{\Upsilon}) \sqrt{ \Upsilon-\eta \sqrt{\Upsilon}} }{ 8 \Upsilon \cos (k \pi/n) } &
 -   \frac{( \Upsilon - \eta \sqrt{\Upsilon}) \sqrt{ \Upsilon+\eta \sqrt{\Upsilon}} }{ 8 \Upsilon \cos (k \pi/n) } \\
\quad & \quad & \quad \\ 
c_2 &  \frac{ \sqrt{-\Upsilon+\eta \sqrt{\Upsilon}} \sqrt{\Upsilon^2-\eta^2 \Upsilon} (-1 + i \tan (k \pi / n) ) }{ 32 \Upsilon \cos (k \pi /n ) } &  
     \frac{-\sqrt{ \Upsilon+\eta \sqrt{\Upsilon}} \sqrt{\Upsilon^2-\eta^2 \Upsilon} ( 1 - i \tan (k \pi / n) ) }{ 32 \Upsilon \cos (k \pi /n ) } \\ 
\quad & \quad & \quad \\ 
c_3 & \frac{ \sqrt{-\Upsilon+\eta \sqrt{\Upsilon}} \sqrt{\Upsilon^2-\eta^2 \Upsilon} (-1 + i \tan (k \pi / n) ) }{ 32 \Upsilon \cos (k \pi /n ) } &
    \frac{ \sqrt{ \Upsilon+\eta \sqrt{\Upsilon}} \sqrt{\Upsilon^2-\eta^2 \Upsilon} ( 1 - i \tan (k \pi / n) ) }{ 32 \Upsilon \cos (k \pi /n ) } \\ 
\quad & \quad & \quad \\ 
c_4 &  \frac{( \Upsilon + \eta \sqrt{\Upsilon}) \sqrt{ \Upsilon-\eta \sqrt{\Upsilon}} }{ 8 \Upsilon \cos (k \pi/n) } &
     \frac{( \Upsilon - \eta \sqrt{\Upsilon}) \sqrt{ \Upsilon+\eta \sqrt{\Upsilon}} }{ 8 \Upsilon \cos (k \pi/n) } \\
\quad & \quad & \quad \\ 
\end{array} \]
\label{eigsyscyclacene}
\end{table*}

\clearpage
\newpage

\section{Appendix III} 

We detail here the reasons that led us to the choice of the imaginary exponential function in order to generalize the position operator to periodic systems.
We limit ourselves to the 1-D case.
These arguments had already been very schematically introduced in Ref. [\onlinecite{Valenca}].
Let us consider the periodic interval (the ``supercell'') $[0,L]$, and let $x$ be the coordinate of a point belonging to the supercell: $x \in [0,L]$.
Let us call $q(x)$ the periodic position associated to the point of coordinate $x$.
We impose the three following general conditions to the periodic position:
\begin{enumerate}
    \item 
    The function $q(x)$ must be a continuous periodic function of period $L$:
    \begin{equation}
        q(x+L) \; = \; q(x) \;, \; \forall x \;  .
    \end{equation}
    In other words, $q(x)$ is translationally invariant in the supercell $[0,L]$.
    \item
    The distance between two points, $x$ and $x+d$, defined as the modulus of the difference between the corresponding complex positions, must be a function of $d$ alone, independent from $x$:
    \begin{equation}
        |q(x+d)-q(x)|^2 \; = \; |q(d)-q(0)|^2 \; .
    \end{equation}
    \item
    For large values of $L$, and $d$ fixed, we must obtain the ordinary distance between the two points:
    \begin{equation}
        \lim_{L \rightarrow \infty} \, |q(d)-q(0)|^2 \; = \; d^2 \; .
    \end{equation}
    In the limit of an infinite supercell, one must recover the non-periodic result.
\end{enumerate}

\hspace{2mm}

Condition 1 is manifestly satisfied choosing for $q(x)$ a function of the type
\begin{equation}
    q(x) \; = \; \sum_{k=-\infty}^\infty a_k \, \exp \Bigl(\frac{i 2\pi kx}{L}\Bigr) \; ,
\end{equation}
with $k$ integer.
In order to investigate Condition 2, we compute the difference $q(x+d) \, - \, q(x)$ by using the previous equation.
We obtain:
\begin{equation}
    q(x+d) \, - \, q(x) \; = \; \sum_{k=-\infty}^{\infty} a_k \, \exp\Bigl(\frac{i 2\pi kx}{L}\Bigr)
    \Bigl[\exp\Bigl(\frac{i 2\pi kd}{L}\Bigr)-1\Bigr] \; .
\end{equation}
We compute now the square of the distance between the points $x+d$ and $x$, given by the square modulus of this quantity, $|q(x+d)-q(x)|^2$.
We get
\begin{equation}
    |q(x+d)-q(x)|^2 \; = \; \sum_{k=-\infty}^{\infty} \sum_{l=-\infty}^{\infty} \, a_k^* a_l \,  \Bigl[\exp\Bigl(\frac{i 2\pi (l-k) x}{L}\Bigr)\Bigr] \, \Bigl[\exp\Bigl(\frac{-i 2\pi kd}{L}\Bigr)-1\Bigr] \, \Bigl[\exp\Bigl(\frac{i 2\pi ld}{L}\Bigr)-1\Bigr] \; .
    \label{eq:dist2}
\end{equation}
Among the three terms within square brackets, the only one containing $x$ is the first exponential factor.
Therefore, in order to have a quantity not depending on $x$, a sufficient condition is that all terms having $l \neq k$ in this equation vanish. 
This happens if only one term in Eq. (\ref{eq:dist2}) survives.
Besides the trivial constant solution $q(x)=a_0$, that does not lead to any physically acceptable result, let us consider a term $a_j$ different form zero. 
One can note that the corresponding $a_{-j}$ term is vanishing.
This fact rules out real solutions of the type $q(x) = a \sin\bigr(\frac{2\pi j x}{L}\bigr)$, {or} $q(x) = a \cos\bigr(\frac{2\pi j x}{L}\bigr)$.
We notice, moreover, that an exponential function is much easier to manipulate than a trigonometric one.
Therefore, Condition 2 suggests the choice, for instance (let us assume $j=1$),
\begin{equation}
    q(x) \; = \; a_1 \, \exp \Bigl(\frac{i 2\pi x}{L}\Bigr) \, + \, a_0 \; .
\end{equation}
It is worth noticing that the presence of the $a_0$ term does not invalidate the request that the quantity in Eq. (\ref{eq:dist2}) does not depend on $x$, given that for $k=0$ or $l=0$ the second or the third term in square brackets is vanishing.
Finally, a Taylor expansion of Condition 3 implies $a_1 \, = \, \frac{L}{2\pi}$.
On the other hand, no physical constraints can be used to fix a value for $a_0$, which is an arbitrary parameter related to the zero of the periodic position.

The above reasons suggest the definition
\begin{equation}
    q(x) \; = \; \frac{L}{2\pi} \, \exp \Bigl(\frac{i 2\pi x}{L}\Bigr) \; ,
\end{equation}
which is the one we use.
The equivalent choice 
\begin{equation}
    q(x) \; = \; \frac{L}{2\pi} \, \exp \Bigl(\frac{-i 2\pi x}{L}\Bigr) \; ,
\end{equation}
is also possible, being simply obtained from the previous one by a parity operation.
The constant term $a_0$ can be chosen equal to $-\frac{L}{2 \pi i}$, in such a way to remove the constant term appearing in the exponential expansion.

Different non-equivalent choices are also possible for the integer $k$, for instance, by choosing a different $a_k$ ($k= \pm 2$, or $k= \pm 3$,...) as the only non-zero term in Eq. (\ref{eq:dist2}) .
In the limit of large boxes, all these choices lead to the same results and are therefore equivalent.
However, the choice of $a_1$ (or equivalently, $a_{-1}$) are those that converge most quickly to the infinite-size limit and are therefore preferable.
Notice that, as far as we have been able to find, no real solution satisfy all the three Conditions, 1-3.
The characteristic of a complex nature is also shared by the operator $\hat U$ introduced in Resta's formalism.
The periodic position seems to be intrinsically complex.

\clearpage
\newpage

\section*{Data Availability}

The data that supports the findings of this study are available within the article.

\end{document}